\documentclass[amsmath,amssymb,aps,twocolumn,prd]{revtex4}

\usepackage{amssymb,graphicx}
\usepackage{amsmath,amsfonts}
\usepackage{color}

\def\vk{\varkappa}
\newcommand{\be}{\begin{equation}}
\newcommand{\ee}{\end{equation}}
\newcommand{\bea}{\begin{eqnarray}}
\newcommand{\eea}{\end{eqnarray}}
\newcommand{\bg}{\begin{gather}}
\newcommand{\eg}{\end{gather}}
\newcommand{\bseq}{\begin{subequations}}
\newcommand{\eseq}{\end{subequations}}

\begin{document}

\begin{flushright}  
CERN-PH-TH/2013-308
\end{flushright}
\vskip -0.9cm
\title{Prospective constraints on Lorentz violation
from ultrahigh-energy photon detection}

\author{Grigory Rubtsov$^{a,b,c}$, Petr Satunin$^a$ and Sergey
  Sibiryakov$^{a,d,e}$} 
\affiliation{{$^a$}\it Institute for Nuclear Research of the
Russian Academy of Sciences, \\   
      \it  60th October Anniversary Prospect, 7a, 117312
      Moscow, Russia}
\affiliation{{$^b$}\it Faculty of Physics, Moscow State University,
\it Vorobjevy Gory, 119991 Moscow, Russia}
\affiliation{{$^c$} \it 
Novosibirsk State University, Pirogov street 2, 630090 Novosibirsk,
Russia}
\affiliation{{$^d$} \it 
Theory Group, Physics Department, CERN, CH-1211 Geneva 23,
Switzerland}
\affiliation{{$^e$} \it 
FSB/ITP/LPPC, \'Ecole Polytechnique F\'ed\'erale de Lausanne,
CH-1015 Lausanne, Switzerland}

\begin{abstract}
We point out that violation of Lorentz invariance 
affects the interaction of high-energy photons with the Earth's
atmosphere and magnetic field. In certain parameter region this
interaction becomes suppressed and the photons escape observation
passing through the atmosphere without producing air
showers. We argue that a detection of photon-induced air showers with
energies above $10^{19}\,\mbox{eV}$, implying the absence of
suppression as well as the absence of photon decay, will put tight
double-sided limits 
on Lorentz
violation in the sector of quantum electrodynamics. These constraints
will be by several orders of magnitude stronger than the existing ones
and will be robust against any assumptions about the astrophysical
origin of the detected photons.
\end{abstract}

\maketitle

\section{Introduction}
One of important open questions in astroparticle physics is the
presence of a photon component in the ultra-high-energy
cosmic rays (UHECR). An answer to it will help to discriminate between
different hypotheses about UHECR production and composition.
The recent progress in the UHECR observations
has revealed a cutoff in the cosmic ray spectrum at 
energies $\sim 5\cdot 10^{19}\,\mathrm{eV}$ 
\cite{Abbasi:2007sv,Abraham:2008ru,AbuZayyad:2012ru}.
This is consistent with the prediction made long ago by 
Greisen, Zatsepin and Kuzmin (GZK)
\cite{Greisen:1966jv,Zatsepin:1966jv} that the flux of primary protons
would be depleted at
these energies
due to pion production on the
cosmic microwave background (CMB). Neutral pions subsequently decay
into photons. Thus, if the GZK nature of the observed cutoff is confirmed,
the fraction of photons in the cosmic rays 
 with energies
$10^{19}\div 10^{20}\,\mathrm{eV}$ is predicted at the level 
$10^{-4}$ -- $10^{-2}$ depending on the
unknown radio background that affects the propagation of photons
through the interstellar medium~\cite{Gelmini:2005wu}. 
If the cosmic ray primaries are nuclei instead of photons,
the cut-off is the
result of nuclei' photodisintegration on the cosmic infrared background
and CMB~\cite{Puget:1976nz} and the predicted 
UHE photon flux is two orders of magnitude
lower~\cite{Gelmini:2005wu,Hooper:2007}. 
UHE photons may also come directly from nearby astrophysical transient
sources \cite{Murase:2009ah}. They also appear in the `top-down'
models of UHECR production, such as superheavy dark matter decay or
Z-burst~\cite{Gelmini:2005wu}. The photon flux predicted by the latter
models is actually higher than the existing bounds, so these models
are already disfavored as the primary source of UHECR
\cite{A+Y,Auger_hyblim2}. Still, they may be allowed at the level of a
subdominant contribution. 

No UHE photon detection has been reported so far. However, the upper
bounds on the UHE photon flux obtained by the current
experiments~\cite{AGASA_Risse,Ylim,Ylim18,A+Y,Auger_sdlim,Auger_hyblim2,TAglim}
are getting close to the value predicted from the GZK process and one
may expect observation of UHE photons in 
the near
future~\cite{Alvarez-Muniz:2013mfa}.

The physics of cosmic rays is known to be highly sensitive
to possible departures from Lorentz invariance (LI) at high energies
\cite{Coleman:1997xq,Jacobson:2002hd}. The motivations for such
departures, mainly rooted in models of
quantum gravity, and their implications for various branches of
physics have been extensively discussed in the literature, 
see the recent review \cite{Liberati:2013xla} and references
therein. The effects 
of Lorentz
violation (LV) are conveniently parameterized within the effective
field theory framework \cite{Colladay:1998fq,Kostelecky:2008ts} that assumes
existence of a preferred frame, commonly identified with the
rest-frame of the
CMB. In this frame UHECR particles have a huge energy which may lead
to the enhancement of LV 
and to observable deviations from the standard physics.

It has been proposed~\cite{Galaverni:2007tq,Maccione:2008iw}
that a detection of UHE photon flux compatible with the GZK prediction
will impose strong constraints on LV in the sector of quantum
electrodynamics (QED). Assuming that the primary cosmic
rays are extragalactic protons and that all UHE photons come from the
GZK process, these papers simulate the propagation of UHE photons
through the interstellar medium and their resulting 
flux on Earth in a
family of LV models.
They showed that, depending on the region in the LV parameter space,
the flux will be either highly suppressed due to the vacuum decay of
photons into $e^+e^-$; or, on the
contrary, it will be enhanced above the existing bounds due to the
weakening of the UHE photon depletion through pair-production 
on the CMB. Similar arguments were used earlier in \cite{Stecker:2001vb} to
set the constraints on LV in QED from observations of TeV gamma rays.

In this paper we point out that not only the photon propagation, but
also its interactions with the Earth's atmosphere and magnetic field
are sensitive to LV.
This effect must be taken into account in the analysis of the
experimental signatures of LV. Due to it the detection of several
photons with energies $10^{19}\div 10^{20}$ eV compatible with the 
standard signatures
will allow to put very restrictive constraints on LV parameters --- at
least an order of magnitude stronger than
in~\cite{Galaverni:2007tq,Maccione:2008iw}. Importantly, these
constraints will be robust against any assumptions about the origin
and propagation of UHE photons towards the Earth. The idea to use
interactions with the atmosphere for constraints on LV was first
proposed in \cite{Antonov:2001xh} in the case of hadronic UHECR 
primaries. It was discussed in the context of TeV gamma rays in
\cite{Vankov:2002gt}.    
 
\section{Interaction of photons with the atmosphere and Lorentz violation}

In the standard LI picture, a primary UHE photon reaching the
Earth interacts in the atmosphere and produces an extensive air shower
of particles with lower energies that can be detected by the
ground-based experiments. The characteristics of the shower are
sensitive to the altitude, at which the first interaction initiating 
the shower occurs. This, in its turn, is determined by
the cross section of the first interaction. 
At energies $\sim 10^{19}\,\mbox{eV}$
the dominant channel of the first interaction is $e^+e^-$ 
pair production on nuclei in the
atmosphere --- the Bethe-Heitler (BH) process. At
higher energy ($10^{20}\,\mbox{eV}$ and above) the pair production in
the geomagnetic field becomes important leading to the formation of a
preshower above the atmosphere. 
The photon showers initiated by these
processes can be identified by the cosmic
ray detectors using various observables \cite{RisseRev}.
As we now discuss,
the cross section of the first interaction, and consequently the
characteristics of the shower, are strongly affected by LV.

Consider the BH process. The standard result for the cross section reads,
\be
\label{BH}
\sigma_{\mathrm{BH}}=\frac{28Z^2\alpha^3}{9m^2}
\Big(\log \frac{183}{Z^{1/3}}-\frac{1}{42}\Big)\;,
\ee  
where $m$ is the electron mass, $\alpha$ is the fine structure
constant and $Z$ is the nucleus  
charge; for scattering on nitrogen ($Z=7$) this gives $\sigma_{BH}\approx
0.51\,\mathrm{b}$. 
One observes that, up to factors of order one, the formula (\ref{BH})
follows from a simple dimensional analysis. The mass $m$ in the
denominator appears because it characterizes the momentum transfer in
the process, while the numerator is obtained by multiplying the
coupling constants at the vertices of the corresponding Feynmann
diagrams.
Depending on the photon energy and the density at the point of
  the first interaction, destructive interference between several
  scattering centers 
  somewhat
suppresses the BH cross section --- the 
Landau--Pomeranchuk--Migdal (LPM) effect. The maximal suppression
occurs for the final configuration with equal energies of electron and
positron and is about a
factor of two \cite{RisseRev}.

The generic effect of LV is the modification of the dispersion relations of
photons and electrons/positrons. At momenta smaller than the 
scale of LV, which is commonly identified with the Planck mass
$M=10^{19}\mathrm{GeV}$, these can be expanded in the powers of 
momenta
\begin{equation}
\label{disprels}
E^2_\gamma=k^2+\sum_{n\geq 3}\frac{a_n k^n}{M^{n-2}}, 
\qquad   E^2_{e^\mp}=m^2+p^2 + \sum_{n\geq 2}\frac{b_n^{\mp}p^n}{M^{n-2}}\;.
\end{equation}
Note that we normalize the low-energy velocity of photons
  to 1, so the quadratic correction to the photon dispersion relation
  is absent.
To get insight about the role of the additional terms one notices that
they can be considered
as effective
momentum-dependent masses of the particles,
\be
\label{meffs}
m_{\gamma,\mathrm{eff}}^2(k)=\sum_{n\geq 3}\frac{a_n k^n}{M^{n-2}},~~~~
m_{e^\mp,\mathrm{eff}}^2(p)=m^2+\sum_{n\geq 2}\frac{b_n^{\mp}p^n}{M^{n-2}}\,.
\ee
The presence of these masses changes
the kinematics of various reactions. In particular, for 
$m_{\gamma,\mathrm{eff}}$ larger than the sum of effective masses of
electron and positron the photon decay in vacuum becomes kinematically
allowed. In this case UHE photons decay almost
instantaneously 
into
$e^+e^-$ pairs and do not reach the Earth
\cite{Stecker:2001vb,Jacobson:2002hd}. 
We focus on the opposite situation, $m_{e^\mp,\mathrm{eff}}\gtrsim
m_{\gamma,\mathrm{eff}}$. 
It is straightforward to see that the characteristic momentum transfer
in the BH process is now given by the overall scale of the effective
masses of particles, which we denote $m_\mathrm{eff}$, evaluated at
the momentum of the incoming photon. The dimensional analysis then
yields an estimate of the cross section,   
\be
\label{BHestim1}
\sigma_\mathrm{BH}^\mathrm{LV}\sim \frac{Z^2\alpha^3}{m^2_{\mathrm{eff}}(k)}\;.
\ee
This is consistent with the estimate of Ref.~\cite{Vankov:2002gt}
obtained for the special case of cubic photon dispersion relation
using the concept of the radiation
formation length and is confirmed by explicit calculation in a model
of LV QED \cite{Rubtsov:2012kb}. 
From (\ref{BHestim1}) we see that 
if the effective mass
evaluated at the energy of the primary photon significantly exceeds
$m$,
the BH
process will be  strongly suppressed. Non-observation of such
suppression will allow to put constraints on $m^2_{\mathrm{eff}}(k)$
at $k\sim 10^{19} \,\mathrm{GeV}$ which will translate into the bounds on
the coefficients in the modified dispersion relations.

Similar reasoning carries over to the case of the photon decay in
the Earth magnetic field responsible for the generation of the preshower,
with the difference that the sensitivity to the effective
mass in this case is exponential~\cite{Satunin:2013an}. 

\section{Prospective constraints}

While the above arguments are very general and apply to a wide class
of LV extensions of QED, to make the quantitative predictions we
focus on a specific model studied in \cite{Rubtsov:2012kb}.
The Lagrangian in the preferred frame reads:  
\begin{equation}
\begin{split}
\mathcal{L}=&\bar{\psi}\left( i\gamma^\mu D_\mu - m \right) \psi -\frac{1}{4}F_{\mu\nu}F^{\mu\nu} 
+\\
&+i\varkappa \bar{\psi} \gamma^i D_i\psi + \frac{ig}{M^2}D_j\bar{\psi}\gamma^iD_iD_j\psi+ \frac{\xi}{4M^2} F_{kj}\partial_i^2 F^{kj}, 
\label{Lagr} 
\end{split}
\end{equation} 
where $\varkappa,\,g$ and $\xi$ are dimensionless parameters, the 
covariant derivative $D_\mu$ is defined in the standard way, 
$D_\mu\psi=(\partial_\mu+ieA_\mu)\psi$. 
Greek indices run from 0 to 3 and are raised and lowered
with the Minkowski metric, while the Latin indices take values 1,2,3
and stand for the spatial components; 
summation over repeated indices is
understood. This Lagrangian contains LV operators
of dimension up to 6 that are  
rotationally invariant in the
preferred frame, gauge invariant, CPT- and P-even. The motivation for
restricting to these terms is 
discussed in
detail in \cite{Rubtsov:2012kb}.
From (\ref{Lagr}) one obtains the dispersion relations for photons
and electrons/positrons of the form (\ref{disprels}) with $a_4=\xi$,
$b^{\mp}_2=2\vk$, $b_4^{\mp}=2g$ and all other coefficients
vanishing. We do not consider the cubic modifications of dispersion
relations as they have already been strongly constrained by other
types of observations \cite{Maccione:2007yc}; from the theoretical
perspective, they are forbidden by postulating the CPT invariance.

Cross sections of several astrophysically relevant processes following
from the Lagrangian (\ref{Lagr}) have been computed \footnote{The
  computation carefully takes
  into account the modification of the Feynman rules in the model (\ref{Lagr}) 
compared to the LI case.} in Ref.~\cite{Rubtsov:2012kb}. 
For processes with an $e^+e^-$ pair in the final state 
the result is expressed in terms of the 
 combination 
\begin{equation}
\label{omegaLV}
\omega_{\mathrm{LV}}(x)=-\varkappa k  - \frac{gk^3}{4M^2}(1+3x^2)
+\frac{\xi k^3}{2M^2}\;.
\end{equation}
Here $x\in[-1,1]$ is defined via the ratio of the 
momenta 
 of the produced electron and positron projected on the direction of
 the incoming photon,  
$
(p_{e^-}\cdot p_\gamma)/(p_{e^+}\cdot p_\gamma)=(1+x)/(1-x)
$.
A straightforward analysis shows that if 
$\omega_{\mathrm{LV}}(x)$ is larger than $\frac{2m^2}{k(1-x^2)}$ 
for some $x$, vacuum
photon decay becomes
kinematically allowed. Below
we concentrate on the values of $\vk,g,\xi$ when this does not happen.
For negative $\omega_{\mathrm{LV}}(x)<-m^2/k$
LV significantly suppresses the cross section of the BH process. 
In the case $1 \ll k|\omega_{\mathrm{LV}}(1)|/m^2\ll
\alpha^{-4} Z^{-4/3}$ the expression for cross section takes the
form \cite{Rubtsov:2012kb}, 
\begin{equation}
\sigma_{\mathrm{BH}}^{\mathrm{LV}} \simeq
\frac{8Z^2\alpha^3}{3k|\omega_{\mathrm{LV}}(1)|}\log\frac{1}{\alpha
  Z^{1/3}}\cdot \log\frac{k|\omega_{\mathrm{LV}}(1)|}{m^2}\;, 
\end{equation}
which is smaller than (\ref{BH}) 
by a factor
$m^2/k\left|\omega_{\mathrm{LV}}(1)\right|$ (up to logarithm). 
Note that this is consistent with the estimate (\ref{BHestim1}) upon
identifying $k|\omega_\mathrm{LV}(1)|$ as the precise expression for
$m_\mathrm{eff}^2(k)$.   
Unlike
the standard case, the cross section is
peaked at the maximal asymmetry between the momenta of the pair,
$x=\pm 1$, hence the appearance of $\omega_{LV}$ at $x=1$.

A future UHE photon detection by cosmic ray experiments would imply
that, on the one hand, the photon decay is kinematically forbidden,
and on the other hand the cross section of the first interaction is not too much
suppressed compared to the standard expectation: otherwise the photon
would go through the atmosphere without developing a
shower \footnote{Another process that can be responsible for the first
  interaction is the direct photonuclear reaction. However, its cross
section is only $10\, \mbox{mb}$ ($1/50$ of
BH cross section) for photons with 
energy $10^{19}\,\mathrm{eV}$ \cite{RisseRev}.}.
Conservatively, 
we require that the cross section does not differ by
more that an order of magnitude \footnote{The large asymmetry between
  the momenta of the pair in the LV case implies that the energy of
  the leading particle almost does not degrade, which further
  suppresses the development of the shower. One expects
  that this effect will also enhance
  the fluctuations in the depth of the shower maximum.
 Not taking this into
account leaves our bounds conservative.}. This gives the bounds,
\be 
\label{omegaLVineqs}
\omega_{\mathrm{LV}}(x)\lesssim  \frac{2 m^2}{k(1-x^2)}\;,~~~~
-\frac{10m^2}{k}\lesssim\omega_{\mathrm{LV}}(1)
\ee 
Applying these constraints to each term in
(\ref{omegaLV}) separately, we conclude that
a prospective detection of photons with energies 
$k\sim 10^{19}\,\mbox{eV}$ will allow to constrain the LV parameters
at the level \footnote{One can be more precise and ask how  
accidental cancellations between the three terms
  in (\ref{omegaLV}) will affect the bounds. Note that the
  degeneracy between the first and the two other terms is lifted by
  detection of several photons with energies differing by a factor
  of few. Thus the bound on the parameter $\vk$ is robust. As for the
  parameters $g$, $\xi$, a straightforward analysis of
  Eqs.~(\ref{omegaLVineqs}) reveals a degeneracy along the ray
  $\xi=2g$, $g < 0$. Here the prospective constraints get weaker and
  become
\[
-10^{-7}\lesssim \xi-2g\lesssim 10^{-4}\sqrt{-g}\;.
\]
}
\be
\label{BHbounds}
|\varkappa| \lesssim 10^{-25}~;~~~~ |g|, |\xi| \lesssim 10^{-7}\;.
\ee
Note that smaller than one constraints on $g,\xi$ imply
trans-Planckian suppression of the quartic terms in the particle
dispersion relations.

The constraints (\ref{BHbounds}) can be compared with the existing bounds.
The best laboratory constraint on the parameter $\vk$
reads $|\varkappa| < 4\cdot 10^{-15}$ 
\cite{Altschul:2010na}, which is by ten orders of magnitude weaker
than (\ref{BHbounds}). The quartic terms in the photon and electron
dispersion relations are constrained respectively from the timing of
distant gamma sources
\cite{Aharonian:2008kz,HESS:2011aa,Vasileiou:2013vra} 
and the analysis of the  
synchrotron
radiation from the Crab Nebula
\cite{Liberati:2012jf} at the level $|\xi|<10^{16}~,~
|g|<10^5$. Finally, our bounds 
on the parameters $g$, $\xi$ are an order of
magnitude stronger than those discussed in 
\cite{Galaverni:2007tq,Maccione:2008iw}.  
It is worth stressing that the prospective constraints (\ref{BHbounds})
rely only on the known physics of the Earth's atmosphere and are
insensitive to any assumptions about the origin of the UHE
photons \footnote{Cf. Ref.~\cite{Klinkhamer:2010pq} that discusses the
robustness of the constraints on LV following from the absence of the
photon decay.}.

\section{How many photons are needed ?}

Let us be more precise and estimate the minimal number of UHE photon
detections required to obtain the bounds (\ref{BHbounds}). 
We focus on a
primary photon with energy $10^{19}\,\mbox{eV}$ and for simplicity neglect
the LPM effect. 
Let $X_0$ be the depth of the first interaction of the photon in
the atmosphere. This is a random variable 
with the exponential
distribution $dP/dX_0=\langle X_0\rangle^{-1}
\exp\big( -X_0/\langle X_0\rangle\big)$. The mean value
of this distribution is determined by the cross section of the first
interaction, $\langle X_0\rangle=m/\sigma$, where $m$ is the average mass of
the atoms of the air (typically, nitrogen). 
It is
$\langle X_0\rangle\approx 50\,\mbox{g}\,\mbox{cm}^{-2}$ for the
standard BH cross section and increases in the LV case.

The depth of the first interaction for a given shower is not
directly observed. What is measured instead is the depth $X_{max}$ 
where the number
of charged particles in the cascade reaches its
maximum~\cite{RisseRev}. 
This is
shifted with respect to $X_0$ by the length of the shower
development,  
$X_{max} = X_0 + \Delta X$, where $\Delta X$ is also a
random variable, whose statistics can be assumed Gaussian due to the
large number of interactions that lead to the development of the
cascade.  
The mean value of
$X_{max}$ for photon showers in the LI theory is $\langle
X_{max}\rangle \simeq
1000\,\mbox{g}\,\mbox{cm}^{-2}$ (which roughly coincides with the
total vertical depth of the atmosphere) and the fluctuations are
 $\sim 80\,\mbox{g}\,\mbox{cm}^{-2}$
\cite{Abraham:2006ar}.
In our analysis we will assume that the mean value and fluctuations of
$\Delta X$ do not change in the presence of LV. This is justified,
since the secondary interactions in the cascade are less energetic
than the first one and the effect of LV on them is
weaker \footnote{Strictly speaking, 
$\Delta X$ and its fluctuations can increase in LV case, but accounting for 
this effect would make the constraints even stronger.}.

$X_{max}$ for photon showers can be measured
either directly with fluorescence detectors or by surface
detectors using the properties of the shower front
~\cite{PierreAuger:2011aa,Tokuno:2012mi}. The latter technique has an
order of magnitude larger exposure but larger $X_{max}$ uncertainty of
$50\,\mbox{g}\,\mathrm{cm}^{-2}$ compared to
$20\,\mbox{g}\,\mathrm{cm}^{-2}$ for fluorescence detectors. To be
conservative, we use the value of $50\,\mbox{g}\,\mbox{cm}^{-2}$
as an experimental uncertainty for $X_{max}$.

We simulate $X_{max}$ for a set of $N$ photon events using the standard LI
distribution. Each set is compared to the distribution of $X_{max}$ in
the LV model using 
the Kolmogorov-Smirnov test and the values of $\langle X_{max}\rangle$
excluded at $95\%$ and $99\%$ confidence level are found. We repeat
this procedure multiple time and identify $\langle X_{max}\rangle$
excluded at the corresponding confidence levels by more than half of
the simulated sets (i.e. we require the statistical power of our
predictions to be higher than $0.5$). The upper limits on $\langle
X_{max}\rangle$  obtained in this way for various values of $N$ are
shown in 
Table 1.
The corresponding upper bounds on the ratio of cross sections of the
first interaction in the standard and LV cases are also shown.

\begin{table}
\begin{center}
\begin{tabular}{|c|c|c|c|c|}
\hline
~&\multicolumn{2}{|c|}{95\%\,CL }
&\multicolumn{2}{c|}{99\%\,CL }\\
\hline
N & $\langle X_{max}\rangle$, $\mathrm{g}\,\mathrm{cm}^{-2}$ 
& $\sigma_{BH}/\sigma^{LV}_{BH}$ & $\langle X_{max}\rangle$, 
$\mathrm{g}\,\mathrm{cm}^{-2}$ & $\sigma_{BH}/\sigma^{LV}_{BH}$\\
\hline
1 &   -   &     - &     -  &  -        \\
2 & 1 880 & 18.6  &    -  &  -        \\
3 & 1 380 &  8.6  &  1 970 & 20.4   \\
4 & 1 270 &  6.4  &  1 600 & 13.0    \\
5 & 1 225 &  5.5  &  1 490 & 10.8    \\
\hline
\end{tabular}
\end{center}
\label{tbl1}
\caption{
Predicted upper bounds on $\langle X_{max}\rangle$ and the suppression of
  the first interaction for  
 $N$ UHE events identified as photons with energy $10^{19}\,\mbox{eV}$. }
\end{table} 

Note that one photon event does not lead to any constraints. Indeed,
no matter how strongly the first interaction is
suppressed, the photon has the conditional probability $10\%$ to
interact in the first $100\,\mbox{g}\,\mbox{cm}^{-2}$ of the
atmosphere, provided that it interacts at all \footnote{The absolute
  probability of the photon interaction with the atmosphere cannot be
  constrained without an a priori assumption about the UHE
  photon flux, which we want to avoid.}. 
Similarly, two
events cannot provide exclusion at $99\%$ CL as even for a flat
distribution there is always a
$1\%$ chance that they both 
happen accidentally in the first $100\,\mbox{g}\,\mbox{cm}^{-2}$.
Thus at least three detections are required to set meaningful
constraints and the exclusion power rapidly grows with $N$. We see
from the table that already
$N=5$ is enough to constrain the ratio of cross sections to be less
than $10$ at $99\%$ CL yielding the bounds (\ref{BHbounds}) on the LV
parameters.  

\section{Constraints from preshower}

Even stronger constraints on LV will be obtained in the case of 
UHE photon events with energies $\gtrsim 10^{20}$ eV and the
preshower signature. The process of the photon decay in the
geomagnetic field that leads to the preshower formation is
exponentially suppressed until the photon energy reaches a certain
value --- the property that effectively turns it into a threshold
reaction. As shown in \cite{Satunin:2013an}, the suppression exponent, 
and hence the threshold energy, is modified
by LV. Namely, in the model (\ref{Lagr}) 
the photon decay width has the form,   
\begin{equation}
\Gamma \propto \exp\left[-\frac{8m^3}{3k eH\sin\varphi}
\left( 1 - \frac{k\,\omega_{\mathrm{LV}}(0)}{2m^2}\right)^{3/2}\right],
\end{equation}
where $H$ denotes the magnetic field, $\varphi$ is the angle between the
photon momentum and the magnetic field and 
$\omega_{\mathrm{LV}}(0)$ is given by (\ref{omegaLV}).
We see that in the presence of LV with
$\omega_\mathrm{LV}(0)<0$ the suppression
 exponent is enhanced compared to the standard case which shifts the effective
threshold energy above $10^{20}$ eV. Requiring that the shift is not
too large implies \footnote{The bound on the positive values of 
$\omega_\mathrm{LV}$ again
  follows from the absence of vacuum photon decay.} 
$|\omega_{\mathrm{LV}}(0)|\lesssim 2m^2/k$.
Applying this inequality separately to the terms with different
  $k$-dependence in (\ref{omegaLV}),
we conclude that a detection of
UHE photons at $k\sim 10^{20}\,\mbox{eV}$
with the preshower signature will yield the
constraints \footnote{Note that the direction of degeneracy of
    the constraints (\ref{preshbound})
    in
    the $(g,\xi)$ plane is different from that appearing in the case
    of 
the BH process, cf. footnote [45].},
\be
\label{preshbound}
|\varkappa|\lesssim 10^{-28}~;~~~~  |\xi-g/2| \lesssim 10^{-12}\;.
\ee
Similarly to the case of $10^{19}$ eV photons several detections are
required for statistically significant exclusion. Note, however,
that one expects the exclusion power of the preshower events 
to be 
higher as an additional information can be gained from the comparison
of the photon decay probability in the geomagnetic field with the
probability of the BH interaction in the atmosphere. 

\section{Concluding remarks}

Two comments are in order. First, we have
focused in this paper on the case of UHE photons with energies
$10^{19}\div 10^{20}$ eV that can originate in the GZK
process. However, our discussion applies essentially without changes
to photon-induced air showers of lower energies. 
Of course, the obtained bounds on the LV parameters will be
weaker in this case: the bounds on $\vk$ and $g,\xi$ are 
inversely proportional to the second and fourth power of the 
 photon energy respectively.
Still, a detection of $10^{17}$ eV photons is already able to probe 
Planck-suppressed LV. Moreover, interesting constraints on LV can be
derived from the existing data on TeV photon-induced showers. We
leave this study for future.   

Second, our analysis implies that if high energy LV is
present in nature at the level exceeding (\ref{BHbounds}), no standard
photon showers with energies above $10^{19}$ eV will be detected by the
cosmic ray experiments. Instead, if 
the resulting $\langle X_{max}\rangle$ 
significantly exceeds the atmospheric depth, the signatures of 
UHE photons will resemble those of
neutrinos: there will be no preshowers and 
the probability of the first interaction in the atmosphere 
will be almost
independent of the depth. In particular, similarly to neutrinos,
such LV photons would be able
to produce   
deep inclined air showers with
zenith angle close to 90$^\circ$. 
However, they can be still discriminated from neutrinos using the 
Earth-skimming channel \cite{Abraham:2009uy} where LV photons 
should give no signal due to the absence of tau-leptons in the
photon shower.

\paragraph*{Acknowledgments} We thank Maxim Libanov, Stefano Liberati,
Thomas Sotiriou and Sergei Troitsky for discussions.
This work was supported in part by
the Russian Federation Government Grant 11.G34.31.0047,
the Grants of the 
President of Russian Federation NS-5590.2012.2, MK-1170.2013, 
by the RFBR grants
11-02-01528, 12-02-01203, 12-02-91323, 13-02-01293, 13-02-12095,
14-02-31429 and by the
Dynasty Foundation.

\end{document}